\documentclass[]{spie}  

\usepackage{amsmath,amsfonts,amssymb}
\usepackage{multirow}
\usepackage{graphicx}
\usepackage[table,xcdraw]{xcolor}
\usepackage{array}
\usepackage{ragged2e}
\usepackage{wrapfig}
\usepackage{upgreek}
\usepackage{gensymb}
\usepackage{siunitx}
 \usepackage{aas_macros} 
\usepackage[colorlinks=true, allcolors=blue]{hyperref}
\newcolumntype{C}[1]{>{\centering\arraybackslash}p{#1}}
\newcolumntype{P}[1]{>{\RaggedRight\arraybackslash}p{#1}}

\title{Optical design and polarimetric performance of a SmallSat UV polarimeter to study interstellar dust: PUFFINS}

\author[a]{Ramya M Anche}
\author[a,b]{Hyukmo Kang}
\author[a]{Kyle Van Gorkom}
\author[a]{Dan Vargas}
\author[a]{Haeun Chung}
\author[b]{Ellie Spitzer}
\author[b]{Meredith Kupinski}
\author[c]{B-G Andersson}
\author[d]{Geoff Clayton}
\author[a]{Ewan S. Douglas}
\author[e]{Luca Fossati}
\author[a]{Victor Gasho}
\author[e]{Sreejith Aickara Gopinathan}
\author[a]{Erika Hamden}
\author[f]{Thiem Hoang}
\author[a]{Marcus Klupar}
\author[g]{Ryan Lau}
\author[h]{Alexandre Lazarian}
\author[i]{Tram N Le}
\author[b]{Joanna Rosenbluth}
\author[j]{Ambily Suresh}
\author[a]{Carlos J. Vargas}

\affil[a]{Steward Observatory, University of Arizona, 933N Cherry Avenue, Tucson, Arizona, 85721, USA}
\affil[b]{James C. Wyant College of Optical Sciences, 1630 E University Blvd, Tucson, AZ 85721}
\affil[c]{McDonald Observatory, University of Texas at Austin, 2515 Speedway Boulevard, Austin, TX 78712, USA}
\affil[d]{Space Science Institute, Boulder, Colorado 80301, USA}
\affil[e]{IWF, ÖAW, Austrian Academy of Sciences, Schmiedlstraße 68042 Graz, Austria}
\affil[f]{Korea Astronomy and Space Science Institute (KASI), 776, Daejeon, Korea}
\affil[g]{NSF NOIRLab, University of Arizona, 950 N Cherry Ave, Tucson, AZ 85719, USA}
\affil[h]{University of Wisconsin–Madison, 500 Lincoln Dr, Madison, WI 53706, USA}
\affil[i]{Leiden University, Rapenburg 70, 2311 EZ Leiden, Netherlands}
\affil[j]{Silicon Austria Labs GmbH, Sandgasse 348010 Graz, Austria}

\authorinfo{}

\begin{document} 
\maketitle
\begin{abstract}
The Polarimetry in the Ultraviolet to Find Features in INterStellar dust (PUFFINS) is a SmallSat mission concept designed to obtain ultraviolet (UV) spectropolarimetric observations to probe the interstellar dust grain properties and to understand wavelength-dependent extinction and star formation. PUFFINS plans to observe 70 UV bright target stars at varying distances within a 1800-3200Å wavelength range with 0.02\%  polarimetric accuracy. PUFFINS uses a simple telescope design with all reflective optics coated with protected aluminum to enhance reflectivity in the UV. The telescope and the spectropolarimeter, which consists of a Wollaston prism and a half-wave retarder, have been carefully selected to be greater than Technology Readiness Level 6 (TRL6). The telescope is designed to exhibit negligible instrumental polarization and crosstalk, significantly reducing the time needed for polarimetric calibration in orbit. The optimum and careful selection of the target stars will enable PUFFINS to observe an expanded and well-defined sample to test the predictions by interstellar grain alignment theory in the observation phase of 9 months. This paper outlines the details of the optical and optomechanical design and evaluates the polarimetric performance of PUFFINS.
\end{abstract}

\keywords{interstellar dust, optical design, UV polarimetry, instrumental polarization, smallsat}

\section{INTRODUCTION}
\label{sec:intro}  
The interstellar medium (ISM) is a dynamic, multi-phased environment consisting of dust, gas, and cosmic rays permeated by radiation and magnetic fields.  Stars are formed from the ISM and interact with and replenish it through stellar radiation, winds, and mass loss. The dust in the ISM provides a source of extinction, shielding complex molecules from dissociation and a cooling mechanism for molecular clouds. While of critical importance in the evolution of the universe, the dust composition, evolution, and dynamics are still not fully understood, especially for the smallest grains. Ultraviolet (UV) polarimetry provides a unique tool to probe the characteristics, dynamics, and evolution of small grain populations through its sensitivity to grain mineralogy, size, and environmental parameters. Dust-induced polarization is an important tracer of interstellar magnetic fields \cite{hiltner1949a, bga2015b}. A better understanding of the small-grain alignment will help probe the magnetic field more reliably through understanding galactic polarization, which is a foreground to cosmological data. The controversy over the claimed first detection of the CMB B-modes \cite{ade2014, planck2015, mortonson2014} shows the importance of this foreground. An accurate removal of foreground polarization requires a detailed prescription of the dust, including all grain sizes and polarizing characteristics.

Previous UV spectropolarimetric missions-the Wisconsin Ultraviolet Photopolarimetry Experiment (WUPPE) and Hubble Space Telescope/Faint Object Spectrograph (HST/FOS)---have scratched the surface of ISM UV polarimetry (observing a total of 28 lines of sight), providing exciting but scientifically inconclusive results, including enhanced UV polarization called Super-Serkowski (SuSep) towards 25\% of the observed targets and polarization in the 2175~\AA\ extinction feature in 7\% of the targets. In addition, the 20~cm sounding-rocket borne Wide-field Imaging Survey Polarimeter (WISP)\cite{cole1997,nordsieck1999} provided broad-band UV imaging polarimetry through a handful of observations, e.g., of stars in the LMC. The measurements challenged our understanding of the nature of interstellar polarization and the characteristics and dynamics of small interstellar dust grains. However, due to insufficient data, these puzzles remain unsolved. Paradigm-shifting developments in the theory and modeling of interstellar grain alignment and the availability of comprehensive, multi-band surveys for target selection now allow conclusive testing of the origins and theoretical mechanisms of small-grain alignment. 

\begin{figure}[t]
    \centering
    \includegraphics[width=1\linewidth]{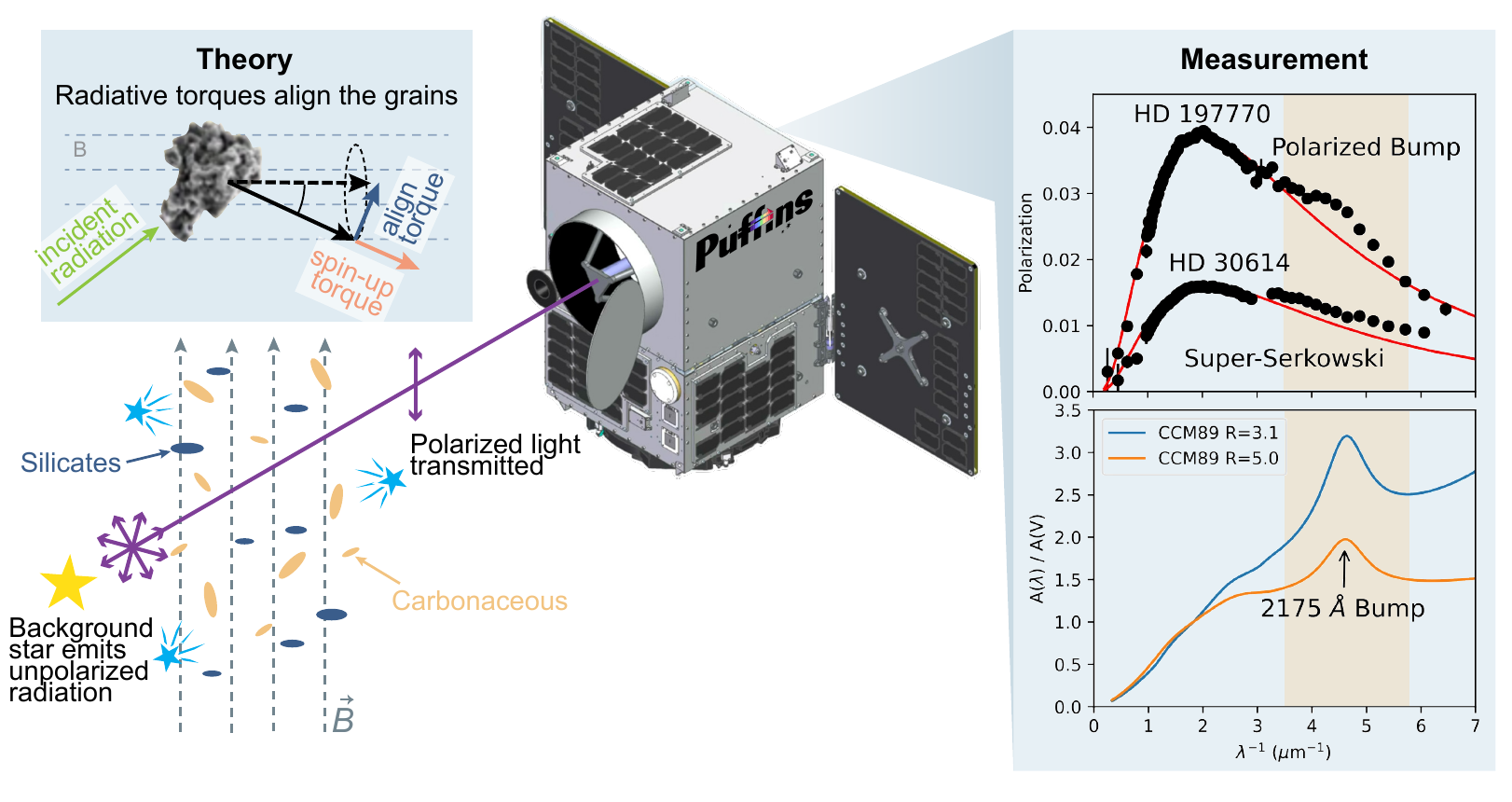}
    \caption{Interstellar polarization is due to the elongated dust grains aligned with a magnetic field. Radiative torques from the differential scattering of the incident radiation's right- and left-hand circular components spin the grains up and align the paramagnetic silicate grains (blue). Carbon solids (orange) are diamagnetic and are not expected to align under most circumstances. The polarization from the small grains will be observed with PUFFINS. The observations will investigate the enhanced UV polarization below 3200 \AA~ and discover the origin of the polarization in the 2175 \AA~ extinction feature. Two measurements made by WUPPE indicating the Super-Serkowski polarization and polarized bump in the UV are shown. Since WUPPE, no UV polarimetry mission has flown, leaving the question of interstellar UV polarization largely unsolved.
    }
    \label{fig:overall-mission}
\end{figure}

In this paper, we present the optical design, analysis and performance of PUFFINS ( shown in Fig.~\ref{fig:overall-mission}) which is a proposed SmallSat UV spectropolarimeter ideal for answering questions related to the dynamics and evolution of the crucial small grain population of ISM. The overall science goals and objectives are discussed in section \ref{sec:science_goals_obj}. Section \ref{sec:science_req} provides the requirements for achieving the science objectives. The optical and opto-mechanical design is explained in section \ref{sec:inst_overview} and estimated instrument performance is presented in section \ref{sec:inst_performance}. The conclusions and discussions are provided in section \ref{sec:conclusion_discussion}. 
\section{Science goals and objectives}
\label{sec:science_goals_obj}
Interstellar dust is pivotal to galaxy evolution, contributing to gas heating, cooling, and star formation. 
Despite constituting just 1\% of the visible mass of the Milky Way, interstellar dust can absorb up to 50\% of UV photons emitted by hot stars. This absorbed energy is reradiated as far-infrared emission. Hence, the properties of interstellar dust grains significantly impact the energy balance within galaxies and influence radiative transfer modeling. The grains tie up varying fractions of the refractory elements and act as surfaces for chemical reactions, from the formation of molecular hydrogen \cite{vandehulst1948,bga2013} to complex organic chemistry \cite{vandishoeck2013,boogert2015}. 
How grains form and evolve, as well as their composition and structure, are long-standing problems of astrophysics. Polarimetry, probing the shape and abundance of aligned and elongated dust grains, is a key diagnostic.

The unifying science goal of PUFFINS is to understand the interactions of the star-forming structures with the ISM by probing the characteristics and dynamics of small interstellar dust grains and how they interact with and trace radiation and magnetic fields. UV spectropolarimetry supported by ground-based data, complemented by extinction curves, provides unique grain shape and mineralogy probes. Through the alignment requirement ($\lambda < 2a$), the UV polarization provides a probe of environments with stellar extreme ultraviolet (EUV) radiation ($\lambda < 912$ \AA).  The first two Science Objectives (SO) of PUFFINS arise from unanswered questions raised by the WUPPE/FOS results and the third SO is motivated by the need for new sampling of selected line of sight with varying characteristics. The three SOs of PUFFINS are as follows:
\begin{itemize}
    \item What causes the enhanced UV polarization seen in $\sim$ 25\% of previous observations? - Investigate the nature and extent of small grain alignment by measuring SuSeP at wavelengths below 3200 \AA
    \item When is the 2175 \AA\ extinction feature polarized and what is its carrier? - Determine the origin and prevalence of the polarization in the 2175 \AA~feature
    \item Survey of ISM UV polarization - Perform a survey of ISM UV polarization of a substantial number of lines-of-sight with varying extinction, metallicity, depletion levels, and ISM/radiation field characteristics, including several targets in each of the Magellanic clouds.
\end{itemize}
\section{Science requirements}
\label{sec:science_req}
\begin{figure}[!ht]
    \centering
    \includegraphics[width=1\linewidth]{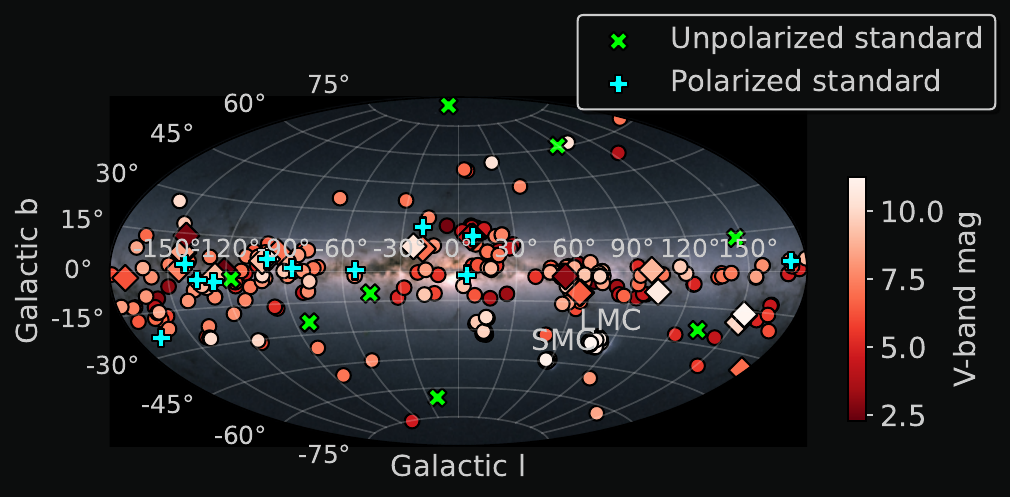}
    \caption{\textbf{A:} Target and calibration star sample for PUFFINS are shown in galactic coordinates. The diamond markers for the target candidates correspond to those stars that have been observed before by WUPPE.}
    \label{fig:science-targets}
\end{figure}

The interstellar dust UV polarization observations require a space-borne telescope for UV sensitivity, but only moderate spectral resolution ($R\gtrsim100$) toward relatively bright stars. The broader science goal for PUFFINS is to measure the wavelength-dependent polarization fraction in the UV for 70 targets carefully chosen to have varying environments, extinctions, metallicities, and depletion levels and are spread across the sky as shown in Figure \ref{fig:science-targets}. 
The science requirements for PUFFINS is provided in Table \ref{tab:science_req}. The field of regard requirement is to access targets across different parts of the sky during the mission lifetime. 
The slit FOV requirement to minimizes background contamination from nearby stars. The lower end of the spectral range ensures some continuum coverage below the polarized 2175 \AA~ features at $\sim$ 1800 \AA. The upper end of the spectral range ensures continuous coverage of the spectropolarimetric curve from the UV to the optical wavelength region when combining the PUFFINS data with ground-based spectropolarimetry. Our spectral resolution requirements  (resolving power) achieve our science objectives of measuring the polarization fraction versus wavelength from 1800-3200 \AA, where the spectral features are broad. Bright blue stars are needed to ensure sufficient flux in the UV down to our short wavelength limit, especially accounting for extinction. 
\begin{table}[!ht]
\centering
\begin{tabular}{|
>{\columncolor[HTML]{DDE8EE}}l |
>{\columncolor[HTML]{CCCCCC}}l |
>{\columncolor[HTML]{EFEFEF}}l |}
\hline
\cellcolor[HTML]{9B9B9B}\textbf{Measurements}                                                                                                                                                                           & \cellcolor[HTML]{9B9B9B}\textbf{Category}                                                       & \cellcolor[HTML]{9B9B9B}\textbf{Requirement}                                                                                                                        \\ \hline
\cellcolor[HTML]{DDE8EE}                                                                                                                                                                                                & Field of regard                                                                                 & All-Sky  Accessible                                                                                                                                                 \\ \cline{2-3} 
\cellcolor[HTML]{DDE8EE}                                                                                                                                                                                                & Single slit FOV                                                                                 & 30   X 30 arcsec2                                                                                                                                                   \\ \cline{2-3} 
\cellcolor[HTML]{DDE8EE}                                                                                                                                                                                                & Spectral   Range                                                                                & 2000   - 3200 Å                                                                                                                                                     \\ \cline{2-3} 
\cellcolor[HTML]{DDE8EE}                                                                                                                                                                                                & Resolving Power                                                                                 & \textgreater   30                                                                                                                                                   \\ \cline{2-3} 
\multirow{-5}{*}{\cellcolor[HTML]{DDE8EE}\begin{tabular}[c]{@{}l@{}}Wavelength-dependent \\ polarization \\ and angle with \\ focus on $\lambda$=   1800-3200Å \\ for 70 unique targets \\ from the target sample\end{tabular}} & \begin{tabular}[c]{@{}l@{}}Polarization   \\ signal to noise ratio \\ (pSNR=p/$\sigma$p)\end{tabular} & \begin{tabular}[c]{@{}l@{}}pSNR   \textgreater 10 for        \\  p (2175 Å)=0.2\% for  \\  a B3 star with V=10 mag \\ for an exposure time of 72 hours\end{tabular} \\ \hline
\end{tabular}
\caption{Science requirements to achieve the science objectives for PUFFINS}
\label{tab:science_req}
\end{table}

\section{Instrument Overview}
\label{sec:inst_overview}
PUFFINS will measure the polarization fraction and position angle from 1800 to 3200 \AA~with a simple Cassegrain telescope and a dual-beam spectropolarimeter. The diameter of the primary mirror (M1) is 250 mm, which is driven by the polarization signal-to-noise ratio (pSNR) requirement for the target sample across all three SOs, as well as the necessity to ensure a sufficient margin on the mission's lifetime.  
The pointing accuracy of 10.0" (1$\sigma$)  drives the size of the slit to 30'' or 320~$\mu$m (with 35\% margin).
The spectropolarimeter employs a well-known and widely used design consisting of a rotatable zero-order achromatic half-wave retarder, a Wollaston prism, and a flat grating, all placed in the collimated beam \cite{schmidt1992} (sec.~\ref{sec:spectropolarimeter}). PUFFINS has been designed to deliver a spatial resolution of 5.4" and spectral resolution $>$ 136 
(sec.~\ref{sec:performance_budget}). The choice of a protective aluminum coating (Al+$\rm \bf MgF_2$) on all mirrors, UV-sensitive scintillator coating on the detector, and a high/optimized grating efficiency provide an average throughput of 11.5\% for unpolarized light. All the subsystems of PUFFINS have been selected to be equal to or greater than Technology Readiness Level 6 (TRL6). PUFFINS builds on the heritage of WUPPE \cite{nordsieck1994} using an $\rm \bf MgF_2$ Wollaston and rotating halfwave $\rm \bf MgF_2$ retarder. PUFFINS has been designed to meet the science requirements in Table.~\ref{tab:science_req} with sufficient performance margins. 
\subsection{Optical design}
The driving science requirement for PUFFINS is to measure a polarization fraction of 0.2\%  with an error of 0.02\% at 2175 \AA. The optical design avoids reflections with large angles of incidence prior to the polarimeter optics to minimize errors due to instrumental polarization. The optical prescriptions of the telescope, guide camera,  and spectropolarimeter are provided in Table.~\ref{tab:telescope_par}, along with key detector parameters. All reflective optics in the spectropolarimeter, including the flat grating, will be coated with Al+$\rm MgF_2$ to achieve a high reflectivity $>$ 85\%.

\begin{figure}[h]
    \centering
    \includegraphics[width=\linewidth]{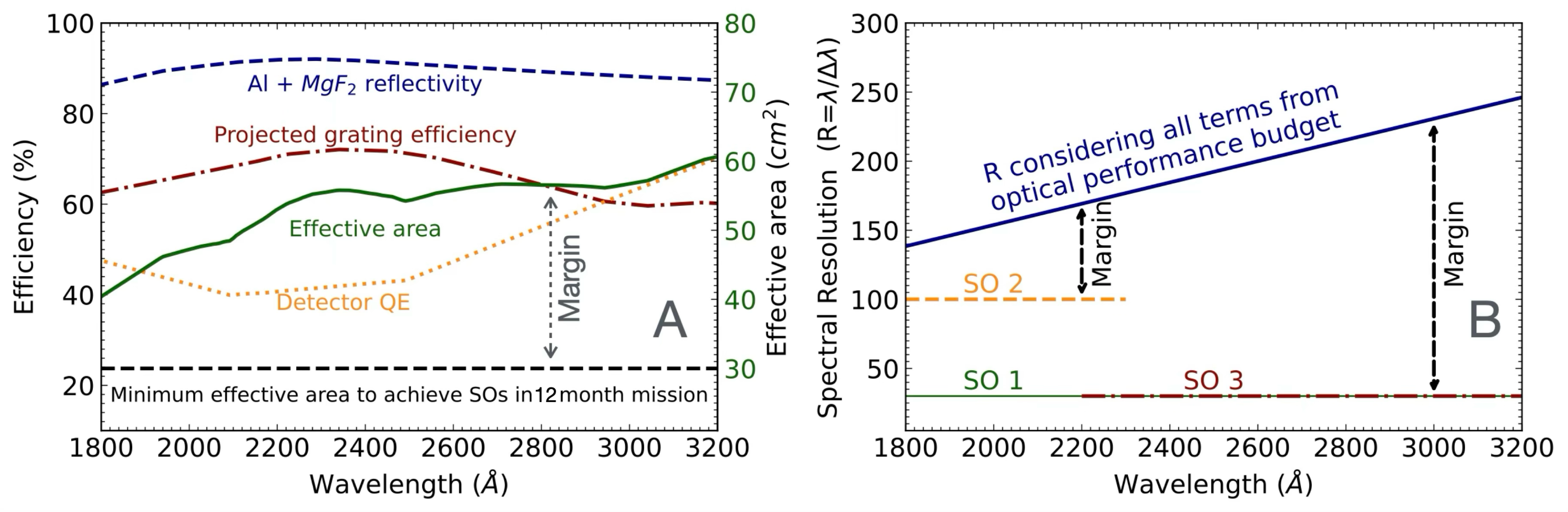}
    \caption{A: Effective area (averaged for both polarizations) and efficiency estimated for PUFFINS. The effective area is sufficient to achieve the SOs in a 8.2-month mission with 46\% margin. B: Spectral resolution and requirements for each of the SOs of PUFFINS as a function of wavelength.}
    \label{fig:effectivearea_spectralresolution}
\end{figure}

The light from the target star is focused on the spectrograph slit, which functions as a field stop for the spectropolarimeter arm, as explained in detail in Sec.~\ref{sec:spectropolarimeter}. The slit mask is tilted to reflect the off-axis stars to a visible wavelength imaging guide camera designed to deliver 1.5'' (1$\sigma$) pointing stability and 2.75'' (1$\sigma$) pointing accuracy. The guider and spectropolarimeter both use a Sony IMX571 CMOS sensor and read out to a SpaceVPX payload computer (TRL6), discussed in Sec.~\ref{sec:detectors}. The effective area curve for PUFFINS is shown in Fig.~\ref{fig:effectivearea_spectralresolution}A, calculated by multiplying the light-collecting area of the M1 (considering the obscuration from the secondary mirror (M2) and the spider structure) with the Al+$\rm MgF_2$ reflectivity (6 reflecting surfaces), the projected grating efficiency from Horiba J-Y (averaged for both the polarizations $\sim$ 55-60\%), and the detector quantum efficiency with the UV-sensitive scintillator coating. The spectral resolution that can be achieved considering all terms from the optical performance budget (Sec.~\ref{sec:performance_budget}) is shown in Fig.~\ref{fig:effectivearea_spectralresolution}B, demonstrating sufficient margin between the requirements and the expected instrument performance. 
\begin{figure}[t]
    \centering
    \includegraphics[width=1\linewidth]{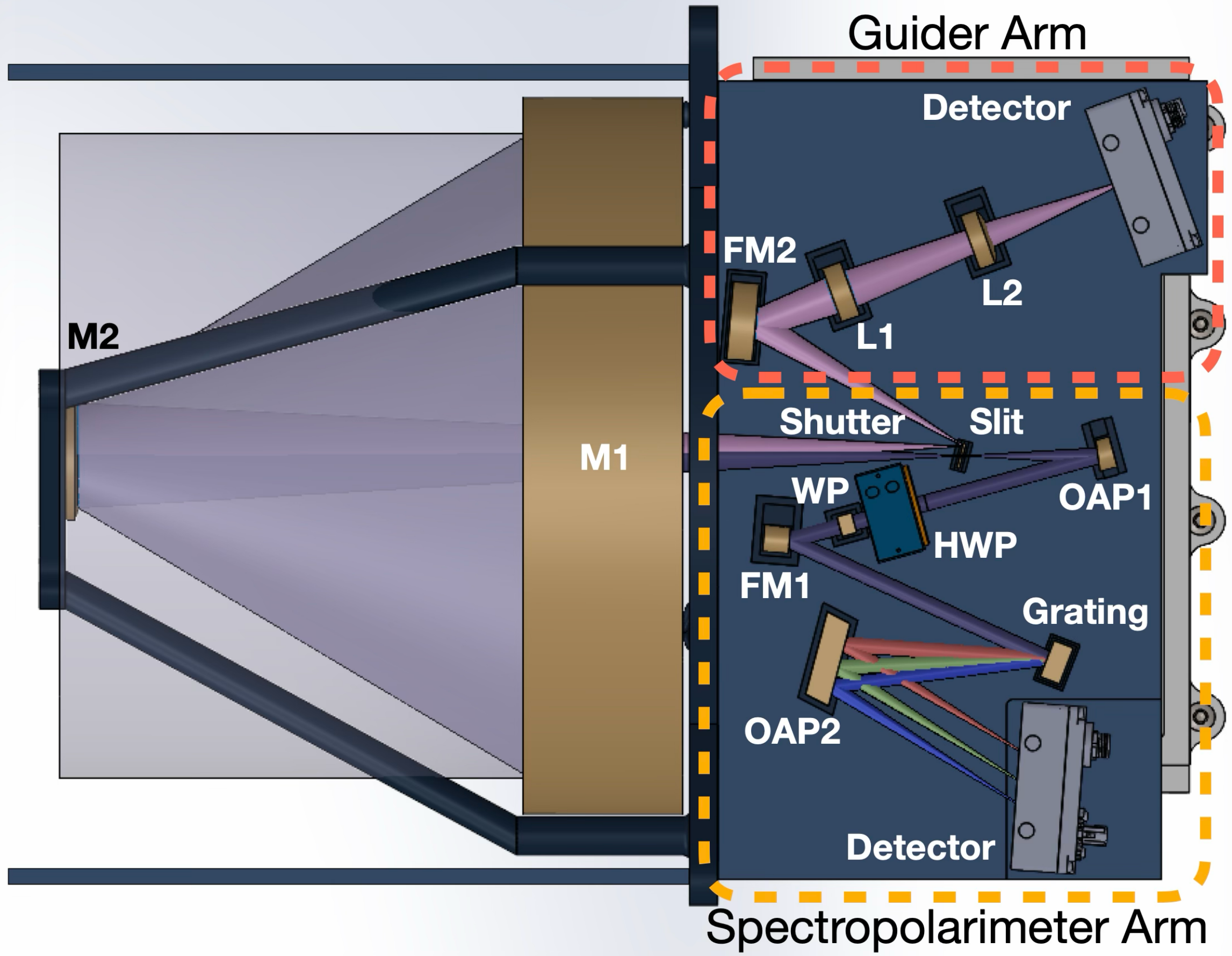}
    \caption{Optomechanical design of PUFFINS}
    \label{fig:optomech}
\end{figure} 
\begin{table}[!ht]
    \begin{minipage}{.5\linewidth}
      \centering
        \begin{tabular}{|P{2.1cm} | P{5.1cm}|}
\hline
\rowcolor[HTML]{3A5D8B} 
\multicolumn{2}{|c|}{\cellcolor[HTML]{3A5D8B}{\color[HTML]{FFFFFF} \textbf{Telescope}}}                \\ \hline
\rowcolor[HTML]{DBE4F0} 
Design type       & Cassegrain                                                                 \\
\rowcolor[HTML]{A5BCD9} 
M1  & \begin{tabular}[c]{@{}l@{}}D=250 mm, RoC=452.46 mm \\
K=-1, Coating: Al+MgF2\end{tabular}   \\
\rowcolor[HTML]{DBE4F0} 
M2  & \begin{tabular}[c]{@{}l@{}}D=50 mm, RoC=81.1 mm \\ K=-1.534, Coating=Al+MgF2\end{tabular} \\
\rowcolor[HTML]{A5BCD9} 
F-number          & 8.5                                                                        \\
\rowcolor[HTML]{DBE4F0} 
Focal length & 2125 mm                                                                    \\ \hline 
\end{tabular} 
\vfill
\vspace{1.6mm}
\begin{tabular}{|P{2.1cm} | P{5.1cm}|}
\hline
\rowcolor[HTML]{F29A79} 
\multicolumn{2}{|c|}{\cellcolor[HTML]{F29A79}\textbf{Guider Arm}}                                           \\ \hline
\rowcolor[HTML]{FDEDE8}\begin{tabular}[c]{@{}l@{}} Fold mirror\\ (FM2) \end{tabular}& \begin{tabular}[c]{@{}l@{}}D=20 mm, \\ Coating: Ag\end{tabular} \\
\rowcolor[HTML]{F6B8A2} 
Lens (L1)                 & \begin{tabular}[c]{@{}l@{}}D=25.4 mm, \\ Achromatic doublet\end{tabular}      \\
\rowcolor[HTML]{FDEDE8} 
Lens (L2)                & \begin{tabular}[c]{@{}l@{}}D=25.4 mm, \\ Achromatic doublet\end{tabular}      \\ \hline
\end{tabular}
\vfill
\vspace{1.6mm}
\begin{tabular}{|P{2.1cm} | P{5.1cm}|}
\hline
\rowcolor[HTML]{CCCCCC}
\multicolumn{2}{|c|}{\cellcolor[HTML]{CCCCCC}{\color[HTML]{000000} \textbf{Detector (Science and Guider)}}} \\ \hline
\rowcolor[HTML]{F2F2F2} 
Sensor             & CMOS, Sony IMX571                        \\
\rowcolor[HTML]{D9D9D9} 
Dark Current       & 0.03 $e^{-}/\mathrm{pix/s}$ at $20^\circ$ C                     \\
\rowcolor[HTML]{F2F2F2} 
Read Noise         & $<$1.5 $e^{-}$ in high gain mode       \\
\rowcolor[HTML]{D9D9D9} 
Coating           & \begin{tabular}[c]{@{}l@{}}UV-sensitive scintillator \\ on science sensor \end{tabular}           \\
\rowcolor[HTML]{F2F2F2} 
QE               & \begin{tabular}[c]{@{}l@{}} 40\% at 2173 \AA, 85\% at 5500 \AA \end{tabular}  \\ \hline
\end{tabular}

    \end{minipage}%
    \begin{minipage}{.45\linewidth}
      \centering
        \begin{tabular}{|P{0.15cm} P{5.5cm}|}
\hline
\rowcolor[HTML]{F5D3A3} 
\multicolumn{2}{|c|}{\cellcolor[HTML]{F5D3A3}\textbf{Spectropolarimeter Arm}}                                                    \\ \hline
\rowcolor[HTML]{FCF4E8} 
\multicolumn{1}{|l|}{\cellcolor[HTML]{FCF4E8}Slit}                                                            & \begin{tabular}[c]{@{}l@{}}On sky=30" $\times$ 30" \\  Size (transmissive)=320$\times$320 $\upmu$m$^2$ \\ Size (reflective) = 7$\times$7 mm$^2$\end{tabular}                                                                                \\
\rowcolor[HTML]{F7DEBA} 
\multicolumn{1}{|l|}{\cellcolor[HTML]{F7DEBA}\begin{tabular}[c]{@{}l@{}}OAP1\end{tabular}} & \begin{tabular}[c]{@{}l@{}}D=12 mm, RoC=104 mm, \\ OAD=14 mm, Coating: Al+MgF2\end{tabular}                                                                                 \\
\rowcolor[HTML]{FCF4E8} 
\multicolumn{1}{|l|}{\cellcolor[HTML]{FCF4E8}\begin{tabular}[c]{@{}l@{}}Modulator\\ (HWP)\end{tabular}}                                                            & \begin{tabular}[c]{@{}l@{}}Achromatic zero-order\\ Half-wave retarder,\\ D=7 mm, thickness=5 mm\\ MgF2, Uncoated\end{tabular}                                   \\
\rowcolor[HTML]{F7DEBA} 
\multicolumn{1}{|l|}{\cellcolor[HTML]{F7DEBA}\begin{tabular}[c]{@{}l@{}}Wollaston\\prism\\ (WP)\end{tabular}}                                                            & \begin{tabular}[c]{@{}l@{}}Size=8$\times$8 mm$^2$, \\ wedge angle=35°, \\ thickness=6 mm, \\ MgF2, Uncoated\end{tabular}                                    \\
\rowcolor[HTML]{FCF4E8} 
\multicolumn{1}{|l|}{\cellcolor[HTML]{FCF4E8}\begin{tabular}[c]{@{}l@{}}Fold mirror\\ (FM1)\end{tabular}}          & \begin{tabular}[c]{@{}l@{}}D=10 mm, \\ Coating: Al+MgF2\end{tabular}                                                                                           \\
\rowcolor[HTML]{F7DEBA} 
\multicolumn{1}{|l|}{\cellcolor[HTML]{F7DEBA}Grating}                                                              & \begin{tabular}[c]{@{}l@{}}Plano reflective, \\ Size=17$\times$15 mm$^2$,\\ Line density: 1300 lines/mm, \\ Blaze angle=8.5$^\circ$,\\ Order=-1, \\ Coating: Al+MgF2\end{tabular} \\
\rowcolor[HTML]{FCF4E8} 
\multicolumn{1}{|l|}{\cellcolor[HTML]{FCF4E8}\begin{tabular}[c]{@{}l@{}}OAP2\end{tabular}}   & \begin{tabular}[c]{@{}l@{}}Size= 20$\times$32 mm$^2$, RoC=165 mm,\\OAD=49 mm, Coating: Al+MgF2\end{tabular}                                                            \\ \hline
\end{tabular}
    \end{minipage}%
     \caption{Telescope and instrument parameters for PUFFINS}
     \label{tab:telescope_par}
\end{table}
\subsubsection{Spectropolarimeter}
\label{sec:spectropolarimeter}
\textbf{Slit:}
The slit functions as a field stop that is designed to block background stars and prevent contamination while maximizing the probability of target acquisition. The slit size is determined by considering i) the distance between the target and background stars and ii) pointing accuracy. The slit is square-shaped with edge length of 320 $\mu$m, equivalent to 30"$\times$ 30" on sky. To enable payload guiding, the slit mask is coated with protected silver and tilted to reflect off-axis (i.e., outside the slit) stars to a visible-wavelength imaging payload guider. A shutter just upstream of the slit will block the beam to both instrument arms. The shutter will be triggered by the spacecraft bus sun sensors for sun protection but is addressable by the payload for calibration activities. 
\begin{figure}
    \centering
\includegraphics[width=0.5\linewidth]{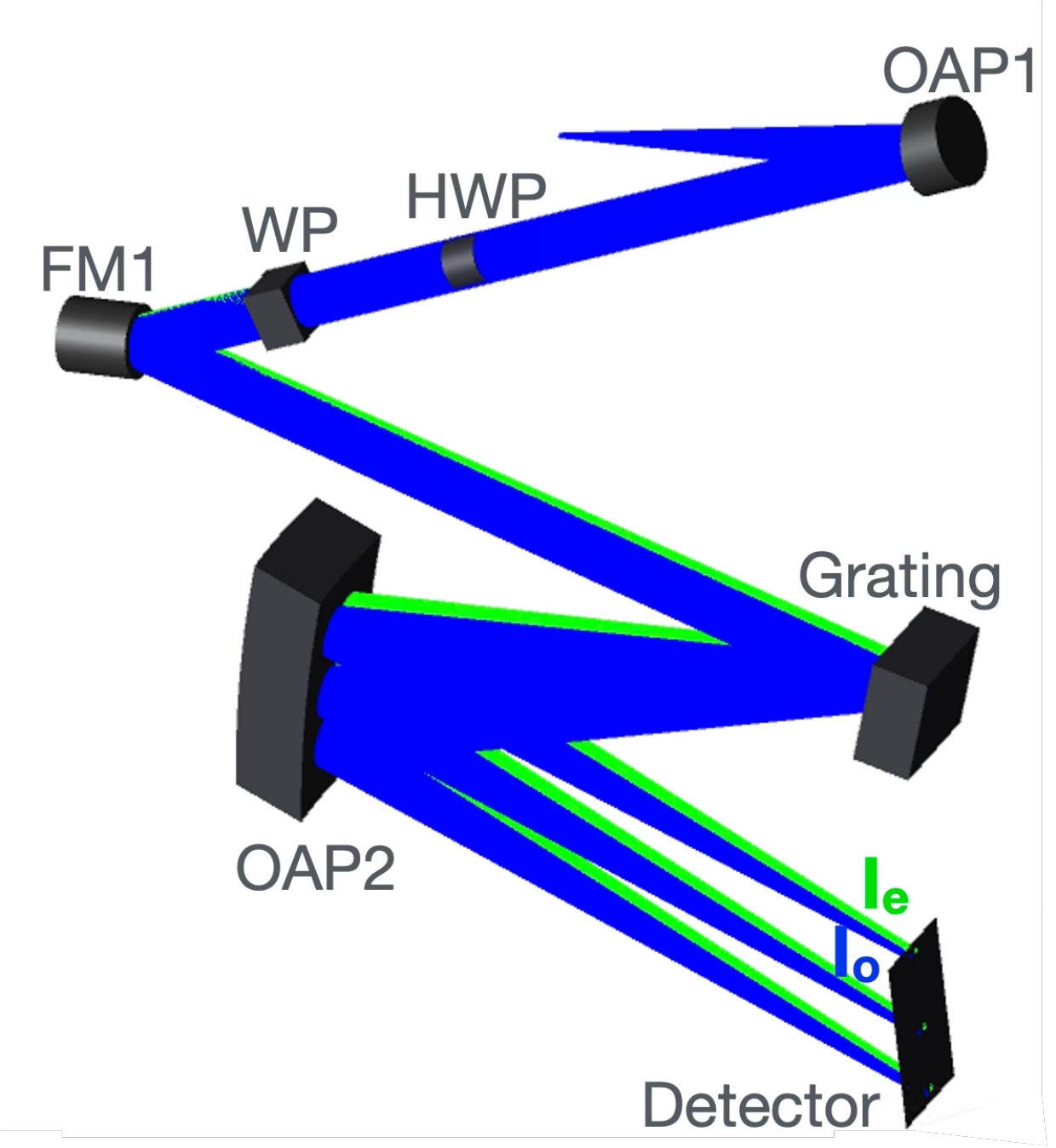}
    \caption{PUFFINS spectropolarimeter optical design}
    \label{fig:opt-design-spec}
\end{figure}

\textbf{Polarimeter:}
The polarimeter has a Wollaston prism as an analyzer and a rotating achromatic zero-order halfwave plate (HWP) as a modulator placed in the collimated beam (after OAP1) to avoid instrumental polarization effects. The Wollaston prism splits the incoming light into two orthogonal polarization states $\rm I_o$ and $\rm I_e$ (see Fig.~\ref{fig:opt-design-spec}). In dual-beam polarimetry, at least two positions of the HWP are required to estimate three unknown parameters: the total intensity ($I$), the polarization fraction $p$, and the position angle of the plane of polarization $\theta$. We need measurements at more than two HWP position because the orthogonally polarized spectra ($I_o$ and $I_e$) fall on different parts of the detector and the polarimetric/optical response of the system is different for each beam ($I_o$ and $I_e$) after the Wollaston prism. PUFFINS will estimate the normalized linear Stokes parameters, $q=Q/I$ and $u=U/I$, from the two orthogonally polarized stellar spectra ($I_o$ and $I_e$) at 8 positions of the HWP between 0\textdegree\ and 180\textdegree\ following a standard measurement procedure in dual-beam polarimetry \cite{serkowski1974polarimeters,keller2002instrumentation,packham2008astrophysical} --- measuring Q at 0\textdegree, 45\textdegree, 90\textdegree, and 135\textdegree and measuring U at 22.5\textdegree,  67.5\textdegree, 117.5\textdegree, and 157.5\textdegree. Thus, each polarimetric observation sequence will include 8 rotations of the HWP, repeated to achieve the required pSNR. Sampling at more HWP rotations additionally allow for tracking of systematic errors, including imperfect optical alignments \cite{patat2006error}. We estimate $q=Q/I$ and $u=U/I$ and corresponding $p$ and $\theta$ as $\sqrt{(q^2+u^2)}$ and $0.5\times \arctan(u/q)$; respectively. The star at the slit location will produce spectra of size 20.27 mm $\times$ 0.27 mm (L$\times$W: 5523 $\times$ 72 pixels) with the two orthogonal polarization beams separated by 1.54 mm (409 pixels) on the detector.

\textbf{Rotation stage for HWP:} 
The HWP in the polarimeter will be rotated using an encoded ultrasonic piezo motor with $3 \upmu$rad resolution rated for ${>} 10$ million revolutions in an ultra-high vacuum (UHV) environment. With 50\% observing efficiency and a $22.5^\circ$ rotation every 10sec, the accumulated revolutions over an eight-month mission lifetime amount to ${<}$ 0.01 million as the upper limit.

\textbf{Grating:} The beams split by the Wollaston prism will undergo dispersion in an orthogonal direction through a reflective diffraction grating. This grating utilizes the -1 diffraction order and features a line density of 1,300 lines/mm on a plano substrate. The design using the plano substrate provides considerable advantages in manufacturing and alignment while meeting the required performance. With a blazing angle of 8.5$^\circ$ and an Al+$\rm MgF_2$ coating, we expect a diffraction efficiency of $>60$\%. 
\subsubsection{Detector}
\label{sec:detectors}
PUFFINS will utilize an integrated sensor and computer package with flight heritage. The science and guide cameras are Sony IMX571 CMOS sensors, which feature $3.76\upmu$m pixels with low read noise ($<$ 1.5 $e^{-}$) and dark current (0.03 $e^{-}/\mathrm{pix/s}$ at $20^\circ$ C). To enhance quantum efficiency (QE) in the UV, a scintillator coating will be applied to the sensor to down-convert incident UV wavelengths to the visible, where detector QE is high. The final QE over the $1800-3200 \AA$  range is expected to be $\gtrapprox 40\%$.
\subsubsection{Payload Guider}
The PUFFINS instrument design includes a payload-integrated guider to deliver 1.5'' (1$\sigma$) pointing stability and 2.75'' (1$\sigma$) pointing accuracy.
The spectrograph slit mask will have a reflective coating and be oriented at an angle to relay off-axis sources to a broadband visible-wavelength guide camera, which will feed back to the spacecraft bus at a 1Hz cadence.
\section{Instrument performance}
\label{sec:inst_performance}
\subsection{Optical error budget}
\label{sec:performance_budget}
The PUFFINS optical error budget is given in Tab.~\ref{tab:optical_budget}. The optical design exclusively employs parabolic and hyperbolic mirrors without freeform optics. This design choice reduces manufacturing costs and simplifies optomechanical alignment.
The most sensitive element in PUFFINS is the M2 of the telescope, particularly with regard to axial translation error. For instance, without focus compensation, the axial translation of M2 must be controlled within $\pm 17\mu$m to meet the spectral resolution requirements. In contrast, the optical elements of the spectropolarimeter are insensitive to misalignments, allowing easy alignment using mechanical methods. The most sensitive element in the spectropolarimeter is OAP1, which induces 0.5" of PSF blur with a 200 $\mu$m of the axial translation error.
The design of PUFFINS, which does not use freeform optics, also has advantages for addressing pointing errors. RMS spot radius varies by only 4.8~$\mu$m across the entire FOV of the slit, even at the longest operational wavelength of 3200 $\AA$. The alignment tolerances for the guide camera are generous, as the effect of misalignment is insignificant on centroid calculations. 
\begin{table}[!ht]
\centering
\begin{tabular}{|P{8.1cm}|P{1cm}|}
\hline
\rowcolor[HTML]{3A5D8B} 
{\color[HTML]{FFFFFF} \textbf{Parameter}} & {\color[HTML]{FFFFFF} \textbf{Value}}\\ \hline
\rowcolor[HTML]{CCCCCC} 
Spectral Resolution Req.           & $\geq$100                                   \\ \hline
\rowcolor[HTML]{CCCCCC} 
Spatial Resolution (1$\sigma$) for \mbox{$R{=}100$} at 1800 \AA & $\leq7.4$"                                     \\ \hline
Optical Design ($1\sigma$)                            & 1"                                     \\ \hline
Surface WFE + Alignment ($1\sigma$)                                 & 2.5"                                     \\ \hline
Pointing Stability ($1\sigma$)                       & 1.5"                                   \\ \hline
Thermal Stability ($1\sigma$)                        & 4.5"                                     \\ \hline
\rowcolor[HTML]{ABC5D4} 
Total Spatial Resolution ($1\sigma$)     & 5.4"                                   \\ \hline
\rowcolor[HTML]{ABC5D4} 
Total Spectral Resolution at 1800 \AA                & 136                                    \\ \hline
\rowcolor[HTML]{ABC5D4} 
Spectral Resolution Margin at 1800 \AA              & 26\%                                   \\ \hline
\end{tabular}
\caption{Optical error budget for spectropolarimeter, which shows a 26\% margin on the spectral resolution requirement at 1800\AA.}
\label{tab:optical_budget}
\end{table}
\subsection{Polarimetric modeling}
\begin{figure}
    \centering
    \includegraphics[width=1\linewidth]{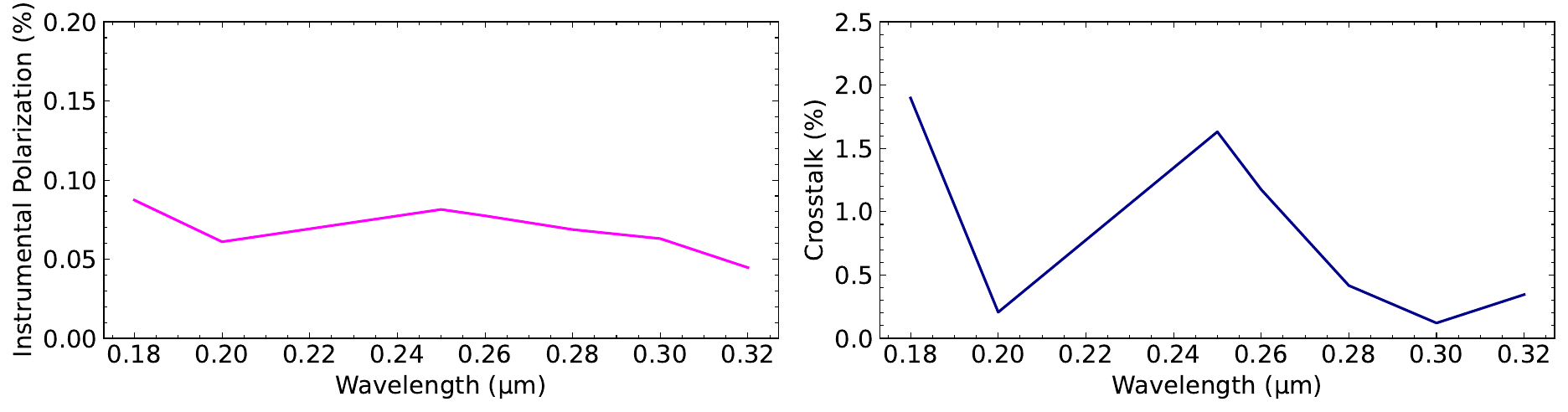}
    \caption{Instrumental polarization and crosstalk estimated before the HWP for PUFFINS}
    \label{fig:ip_ct}
\end{figure}

Instrumental polarization (IP) and crosstalk (CT) arise from Fresnel reflections at the mirror surfaces of the telescope and instrument. The PUFFINS primary and secondary mirrors are designed to be rotationally symmetric to minimize polarization effects. The \ang{9}  angle of incidence on OAP1 results in less than 0.1\%  instrumental polarization and less than 2\% crosstalk before HWP as shown in Figure \ref{fig:ip_ct}. When observing a polarization fraction of 1\%, we expect only 0.02\% of linear to circular crosstalk, which is well below the polarimetric accuracy ($\sigma_p$) of PUFFINS. During ground testing, the spectropolarimeter will be calibrated using a well-established dual rotating polarization state generator (PSG) and polarization state analyzer (PSA) to measure the Mueller matrix of the entire optical system. During on-orbit operations, polarization calibration will be conducted weekly by observing one polarized and one unpolarized standard star from the standards list and that are spread across the sky (shown in Fig.~\ref{fig:science-targets}).
\subsection*{Projected performance}\label{sec:performance}
\begin{figure}[!ht]
     \centering
     \includegraphics[width=0.95\linewidth]{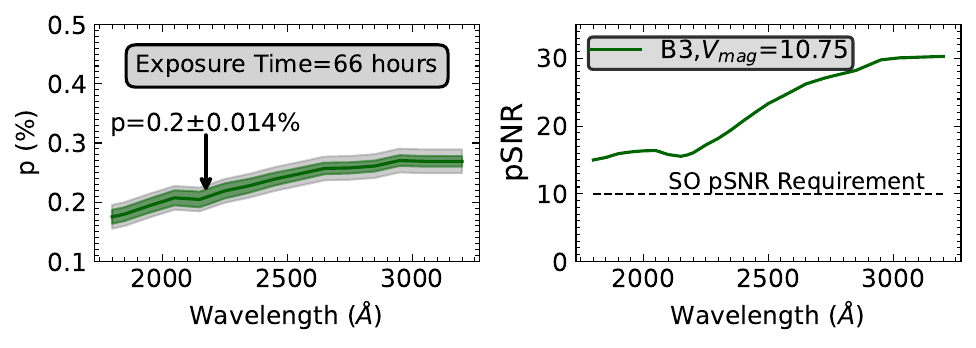}
     \caption{Project p and pSNR performance of PUFFINS for a $V_{mag}$=10.5 of B3 type star meets the requirement in 66 hours of observation.}
    \label{fig:projected_per}
\end{figure}

We developed a SNR calculator for PUFFINS to estimate the polarization fraction and the pSNR for a star of $V_{mag}$=10.5 of B3 spectral type from the targets candidates shown in Fig.~\ref{fig:science-targets}. We input a wavelength dependent polarization curve with polarization fraction (p) varying between 0.14 to 0.27 to the SNR calculator. Using the effective area and noise characteristics of the detector of PUFFINS, we calculate the total number of photons/s. We then convert the total number of photons into orthogonal components: ($I_{0}$, $I_{90}$)  ($I_{45}$, $I_{135}$) for 8 HWP positions as measured by PUFFINS for the total exposure time of 66 hours. We estimate the p, $\sigma$p, and pSNR from these measurements, as explained in Sec.~\ref{sec:spectropolarimeter}. Fig.~\ref{fig:projected_per} shows that for a $V_{mag}$=10.5 of B3 type star, PUFFINS will be able to measure the p=0.2 with pSNR$>$10 in 66 hours meeting the requirements provided in the Table.~\ref{tab:science_req}.

\section{Conclusion and Discussions}
\label{sec:conclusion_discussion}
A full understanding of interstellar UV-IR polarization and testing of theoretical models requires coverage of polarization measurements from the far-IR to the UV. We have presented the science goals, requirements, optical design and performance for a smallsat UV spectropolarimeter, PUFFINS. 
The projected performance shows that PUFFINS will be able to measure a polarization fraction of 0.2 with an accuracy of 0.02 and perform a survey of 70 bright stars in mission lifetime of 9 months with sufficient margin. Combined with ground-based optical spectropolarimetry, PUFFINS observations will address key questions about the interaction between interstellar dust grains and their environment, as well as probe magnetic fields through a UV polarization survey of the ISM. PUFFINS will act as a testbed for HWO science cases, including for possible UV spectropolarimetry instruments (e.g., POLLUX\cite{POLLUX2024}) and enhance its scientific output.

\appendix    

\acknowledgments 

Portions of this research were supported by funding from the University of Arizona Space Institute (UASI).

\bibliography{report,bgbiblio_tot} 
\bibliographystyle{spiebib} 

\end{document}